\documentclass[journal]{IEEEtran}
\usepackage [c3,nocomma]{optidef}
\usepackage{algorithmic,algorithm}

\usepackage{diagbox}
\usepackage{amsmath,amssymb}
\usepackage{enumerate,multirow,subfigure,graphicx,float,capt-of}
\usepackage{algorithmic,algorithm}
\usepackage[parfill]{parskip}
\usepackage[numbers,sort&compress]{natbib}
\usepackage{multirow}
\usepackage{caption}
\usepackage{bm}
\usepackage{tabulary}
\usepackage{multicol}
\usepackage{multirow}
\usepackage{cases}
\usepackage[a4paper, total={6.5in, 8.5in}]{geometry}

\usepackage{enumitem}
\usepackage{gensymb}
\usepackage{graphicx}
\usepackage{float}

\begin{document}
\title{Optimal Tree Topology for a Submarine Cable Network With Constrained Internodal Latency}
\author{Tianjiao~Wang,
        Zengfu Wang,
        Bill Moran,
        Moshe Zukerman, \IEEEmembership{Life Fellow,~IEEE}

\thanks{This work was supported in part by the Research Grants Council of the Hong Kong Special Administrative Region, China under
Project CityU8/CRF/13G, in part by the City University of Hong
Kong under Project 9667193, and in part by the Shenzhen Municipal
Science and Technology Innovation Committee under Project JCYJ20180306171144091. \emph{(Corresponding author: Zengfu Wang.)}}
\thanks{Tianjiao Wang and Moshe Zukerman are with the Department of Electrical Engineering, City University of Hong Kong, Kowloon, Hong Kong (e-mail: tianjwang6-c@my.cityu.edu.hk; m.zu@cityu.edu.hk).}
\thanks{Zengfu Wang is with the Research \& Development Institute of Northwestern Polytechnical University in Shenzhen, Shenzhen 518057, China, and also with the School of Automation, Northwestern Polytechnical University, Xi'an 710072, China. (e-mail: wangzengfu@nwpu.edu.cn).}
\thanks{Bill Moran is with the Department of Electrical and Electronic Engineering, University of Melbourne, Melbourne, VIC 3010, Australia (e-mail: wmoran@unimelb.edu.au).}
}

\maketitle

\begin{abstract}
This paper provides an optimized cable path planning solution for a tree-topology network in an irregular 2D manifold in a 3D Euclidean space, with an application to the planning of submarine cable networks. Our solution method is based on total cost minimization, where the individual cable costs are assumed to be linear to the length of the corresponding submarine cables subject to latency constraints between pairs of nodes. These latency constraints limit the cable length and number of hops between any pair of nodes. Our method combines the Fast Marching Method (FMM) and a new Integer Linear Programming (ILP) formulation for Minimum Spanning Tree (MST) where there are constraints between pairs of nodes. We note that this problem of MST with constraints is NP-complete. Nevertheless, we demonstrate that ILP running time is adequate for the great majority of existing cable systems. For cable systems for which ILP is not able to find the optimal solution within an acceptable time, we propose an alternative heuristic algorithm based on Prim's algorithm. In addition, we apply our FMM/ILP-based algorithm to a real-world cable path planning example and demonstrate that it can effectively find an MST with latency constraints between pairs of nodes.

\end{abstract}

\begin{IEEEkeywords}
Integer Linear Programming, Minimum Spanning Tree, cable path planning, latency constraints.
\end{IEEEkeywords}

\section{Introduction}
\label{sec:Introduction}

We have experienced an explosive growth of internet traffic over the last several decades that is expected to continue with the rapid development of 5G, IoT and AI technologies, especially considering the current COVID-19 outbreak. Cisco's latest report that predates the COVID-19 outbreak states that global annual IP traffic will reach 4.8 ZB per year by 2022~\cite{cisco}. As the COVID-19 pandemic places many countries in lock down and many people are working (and learning) from home, the consumption of internet services increases dramatically. Generally, as the result of the pandemic, internet traffic is 25$\%$ to 30$\%$ higher than usual~\cite{thenewstack}.

Submarine cables form a critical component of the international data transmission system, carrying more than 99$\%$ of global IP traffic~\cite{Submarinereport2018}. As IP traffic is growing larger, the construction of additional submarine cables and their path planning optimization are key for meeting the ever-increasing internet traffic demands and provision of cost-effective and reliable internet services.

An important factor in cable path planning optimization is the cost of cable construction. While the cost may depend on several factors, such as future cost consequences of cable breakage associated with earthquakes or fishing activity, and the legal requirements to avoid certain areas, as discussed in~\cite{wang2019cost}, for this paper, for simplicity, we regard it as a linear function of cable length. That is, the cost of the cable between two nodes is assumed to only be based on the length of the geodesic in an irregular 2D manifold in a 3D Euclidean space. Our simplified assumption will be applicable to areas where the above mentioned factors are not applicable. In Section~\ref{sec:Application of ILP Algorithm} we give an example for a region the Mediterranean, to show that our solution using this assumption is almost identical to a solution based on risk consideration where all of the risk factors are taken into account in the cable path design.

Based on the Fast Marching Method (FMM)~\cite{kimmel1998fast, SethianJournal1999}, we find an optimal cable path between two nodes and its optimal length and cost (linear in the length). Currently, the cost of submarine cable construction is estimated at around 24,000 USD per kilometer indicating  a significant  cost of a long-haul submarine cable that may be in the  tens or even hundreds of millions of dollars. Accordingly, a procedure to find the minimum length and/or cost of laying a cable path network, becomes an important part of constructing a submarine cable system~\cite{burnett2013submarine, Elias}. As mentioned above, in this paper, we focus on the case, where only the cable length affects the cost. Henceforth, we will use the term  {\it FMM -- length only} to refer to this approach.
For comparison, we will also use the approach of \cite{wang2019cost} where the cable path is based on FMM involving a range of considerations such as cable survivability~\cite{wang2017seismic}, and this approach will be called {\it FMM -- length and other considerations}. An even simpler approach than FMM -- length only is the one based on the great-circle distance between any two nodes. We will use the common term of {\it great-circle distance} for this approach. Although the great-circle distance is a good approximation to the path length, it does not provide the actual cable path since it does not consider the geographic terrain.

Another important criterion considered in cable path planning is latency, which is the time it takes for a data packet to travel from the sender to the receiver. Latency includes transmission time as well as propagation and queuing delays. Propagation delay which is linearly proportional to the cable length is a significant source of latency~\cite{ xiao2008technical}.  In addition, for a cable network, some pairs of nodes are not connected directly by a single cable, so the data transmitted between them need to go through other nodes which increase queuing delay and therefore latency. From the report in~\cite{broadcast}, each $1,000$ kilometers of cable length produces approximately $10$ milliseconds round trip delay. Clearly, longer cable length and more intermediate nodes will affect the users'experience of some latency-critical applications and may even inhibit the use of such applications.

One example of such latency-critical applications is an autonomous-vehicle system. As autonomous-vehicle technologies evolve, the data generated by vehicles grow exponentially~\cite{kasture2016tailbench}. The execution of autonomous driving needs to be real-time, and according to~\cite{processing}, the time latency must be lower than 10 ms. Processing this substantial amount of data in a fast and seamless way is one of the main challenges for the development of autonomous vehicles~\cite{processing}. However, the penetration rates of autonomous vehicles may vary from region to region. Therefore, for areas/cities where autonomous driving will be heavily used, strict data transmission latency to the data center is likely to be required.

 Latency has also become an important performance consideration for areas such as cloud computing, finance, and content providers. For finance, in many cases, reduction of latency by few milliseconds can significantly increase trading profitability. This is especially true for high-frequency trading that requires the lowest available latency between trading centers \cite{broadcast,computerweekly}. Low latency also plays important role in improving performance of internet services for users. According to reports from online search companies that include Google and Bing, increased latency adversely affects their businesses because it reduces the number of clicks and internet searches. Research done in Bing has indicated that a two-second slowdown would reduce revenue per user by $4.3\%$. The reduction in latency can improve performance, profitability and increase sales for customers. Amazon revealed that every latency of 100 milliseconds causes reduction of its sales by $1\%$~\cite{computerweekly}.

 Online games also have strict latency requirements. On average, if the latency is increased by 100 ms, players reduce their QoE ratings by 14$\%$~\cite{ manuel2014effects}. It is predicted that the sales of competitive games in 2020 will reach $11$ billion and in reality this may even further increase as more people stay at home because of the COVID-19 pandemic and play games. This is evidenced by reports on increase by 70-75\% in gaming activities in USA and Italy \cite{king2020problematic}. Game developers and publishers are finding methods to ensure that their users receive superior QoE. They plan to build new submarine cables to achieve 1TB-per-second bandwidth~\cite{vbtransform}.

 According to the Submarine Cable Almanac of
 2016 ~\cite{SubmarineTeleForum2020}, of all 266  submarine cable systems in the world, 246 have a tree topology (152 out of these 246 systems are point-to-point topology, and the remaining 94 cable systems use trunk-and-branch  topology). In 36 cable systems that are now in the planning stage, 12 of them use point-to-point topology and 12 of them use trunk-and-branch topology. A tree topology (including both point-to-point and trunk-and-branch) is the most commonly used topology in submarine cable systems.


In this paper, we aim to limit the time latency between pairs of nodes according to their requirements while minimizing the overall construction cost of the cable network. Nodes with strict latency requirements are either located near data centers, or they are heavy users of latency-critical applications that require limited latency in their communications with data centers. The contributions of this paper are as follows.

We propose a new perspective to optimize cable network planning. We regard minimizing cost of cable network problem as a Minimum Spanning Tree (MST) problem and consider the latency constraints between pairs of nodes. To our best knowledge, latency constraints between pairs of nodes have not been considered in the research of cable network planning.

We provide, for the first time, a new method for the MST problem over an irregular 2D manifold in 3D space with constraints that include the length as well as the number of intermediate nodes between pairs of nodes. Our new method combines the FMM -- length only which is used to find the optimized cable path between pairs of two nodes and its cost, as well as a new Integer Linear Programming (ILP) formulation that provides a tree-topology cable network at minimal cost and also satisfies the latency constraints for any pair of nodes. We analyze the number of the variables and constraints in the formulation and illustrate the complexity of our ILP-based algorithm via run-times with graphs with a various number of nodes and constraint requirements. For large-scale cable systems for which the ILP-based algorithm is not able to find the optimal solution within the time limit, we propose an alternative heuristic algorithm based on Prim's algorithm.

The remainder of this paper is organized as follows. In section~\ref{sec:rel_res}, we review related research on cable path planning. In Section~\ref{sec:prob_form}, we model the problem of a submarine cable network with constrained latency between pairs of nodes as an MST with constraints problem. In Section~\ref{sec:ILP Algorithm for Minimum Spanning Tree} we propose an ILP formulation to solve this problem for wide majority scale of existing cable systems, and an alternative heuristic algorithm for very few large-scale cable systems. A real-world example that uses our method to achieve optimal cable network planning is shown in Section~\ref{sec:Application of ILP Algorithm}. Finally, Section~\ref{sec:conclusion} concludes this paper.

\section{Related Work}
\label{sec:rel_res}
Many research publications on cable path planning focus on minimizing the total cable cost under the survivability constraints~\cite{agrawal2019network,cao2013survivable,Zhao2016Journal, msongaleli2016disaster,tran2016multi,wang2019cost, wang2019path}. Msongaleli {\it et al.}~\cite{ msongaleli2016disaster} considered a set of possible routes between nodes and a set of disaster scenarios with a probability model for cable break  and provided an ILP-based algorithm to design a submarine cable network to minimize the expected cost in case of a disaster. In~\cite{Zhao2016Journal}, Zhao {\it et al.} proposed a path planning method that aims to obtain a path aiming to minimize cable cost and earthquake risk to the cable using a semi-supervised model based on raster graphics. They used the Dijkstra's algorithm to minimize cost and cable break risk as a result of an earthquake. In~\cite{tran2016multi,wang2019cost, wang2019path}, the approach we call FMM -- length and other considerations was used to provide a solution for a multi-objective (cost and risk) path-planning problem on a 2D manifold in a 3D space that models the earth's surface. In addition to considering presence of earthquake-activities, various other design considerations (such as water depth, sediment hardness and human activities) were considered in the cable path design.

Most of these existing work on cable path planning so far focused mainly on point-to-point path optimization. Except for point-to-point cable design, there is still remains the problem of choosing the optimal topology for the  cable system, and as mentioned above, a tree is a widely used topology for cable systems.  The paper closest to the present paper in terms of the cable path planning is~\cite{ wang2020optimal} which optimizes the cable network in a tree topology. In~\cite{ wang2020optimal}, Wang {\it et al.} considered  path planning for a cable system with a trunk-and-branch tree topology on the earth's surface and formulated the problem as a Steiner Minimal Tree problem. Tran {\it et al.}~\cite{ tran2016enhancing} presented a dynamic programming method to choose new links and routes for a given network. None of these cable path planning methods considered latency constraints for different pairs of nodes which is the main contribution of this paper.

The MST problem is one of the most typical combinatorial optimization problems, and it is heavily considered in this paper. Related work on the MST problem includes work on the Kruskal's algorithm which generates forests in the process of obtaining the MST and find an edge of the least possible weight that connects any two trees in the forests~\cite{kruskal1956shortest}. The computational complexity of Kruskal's algorithm is $O(E * \log(N))$, where $E$ and $N$ represent the number of edges and nodes in the graph, respectively. Prim's algorithm is slightly different from Kruskal's algorithm. It starts from a node $n_0$, then selects the least-cost edge $e (n_0, n_x)$  and adds it and the node $n_x$ to the spanning tree. The computational complexity of the Prim's algorithm is $O(N^2)$. Kruskal's and the Prim's algorithms can also be used to find the MST with constraints by modification. In~\cite{oh1997constructing}, Jaewon and Pedram proposed a novel algorithm, named bounded path length Kruskal (BKRUS), to find MST with constraints. They assumed there is a central node in the graph. The objective is to find an MST that satisfies the distance constraint $D$ from every other node to the central node. However, BKRUS cannot ensure that other pairs of nodes satisfy distance constraints.

 Over the last couple of decades, as a result of advances in high-performance computing and more efficient algorithms, ILP has achieved considerable success in obtaining optimal solutions to many combinatorial optimization problems. Existing ILP formulations for the MST problem include  Martin's Formulation, the Subtour Elimination Formulation, the Cutset Formulation~\cite{martin1991using, fan2014integer,dara2019integer}.

\section{ Problem Formulation and Modeling }
\label{sec:prob_form}
In this paper, our objective is to construct a tree-topology cable network without additional Steiner nodes at minimal cost, while satisfying constraints on certain paths between specified nodes, possibly involving more than one hop between those  nodes. To this end, we consider the cable network as a spanning tree (in the sense of graph theory) in which the cost of every edge is based on the length of the geodesic between its end-nodes. Such costs can be calculated using FMM -- length only. Certain pairs of nodes need different requirements of latency, as discussed in Section~\ref{sec:Introduction}, where the constraints represent a number of possible considerations including physical distance (cable length) or the number of intermediate nodes. In particular, while focusing on minimizing the cost of the entire cable network system, we ensure that cable length and number of intermediate nodes between any pair of nodes satisfy given (achievable) latency requirements.

Let $\mathbb{D}$ be a closed and bounded path-connected region on the surface of the earth where we aim to lay the cable network, and $1, 2, \ldots, n \in \mathbb{D}$ denote the nodes to be connected in a cable network with spanning tree topology. Let $V = \{1, 2, \ldots, n\}$. Let $(i,j),  i, j \in V $ denote the edge connects the nodes $i$ and $j$ on $\mathbb{D}$. Let $E$ denote the set of edges, and $G = (V, E)$ is the graph.
Let $l_{i,j}$ denote the length of the cable path between nodes $i$ and $j$ which may include one or several individual cable edges. Let $N_{i,j}$ denote the number of intermediate nodes between nodes $i$ and $j$, and $\mathbb{C} = \{(i \leftrightarrow j ), i,j \in V \}$ denote the set of pairs of nodes with constraints.

A spanning tree $T$ of $G$ is a connected subgraph which does not contain any cycles. That is, $T=(V, E^*)$, $E^* \subseteq E$ and $T$ is a tree. Denote the total cost of $T$ by $c(T)$.
We aim to find the $T$ with minimal total cost and meanwhile satisfies the constraints given by $\mathbb{C}$. We formulate the problem in the following equations:
\begin{mini!}
	{T}
	{\sum c(T), \label{eq:PMSTlpmin}}
	{\label{eq:PMSTlpmin13}}{}
	\addConstraint{l_{i,j}}{\leq l^*_{i,j}, \,\, \forall \, (i \leftrightarrow j) \in \mathbb{C}, \label{eq:PMSTlpst1}}
	\addConstraint{N_{i,j}}{\leq N^*_{i,j}, \,\, \forall \, (i \leftrightarrow j) \in \mathbb{C}, \label{eq:PMSTlpst2}}
\end{mini!}
where $l^*_{i,j}$ is the corresponding threshold for the length constraint, and $N^*_{i,j}$ is the corresponding threshold for the nodes number constraint.

\section{Solution Methodologies}
\label{sec:ILP Algorithm for Minimum Spanning Tree}
In this section, we first formulate a new ILP for our problem based on Martin's formulation. We assess its computational burden by calculating its number of decision variables (namely, $x_{ij},y_{ij},z^{ab}_{ij}$) and constraints and note that it is NP-complete.
In addition, we propose a heuristic algorithm based on Prim's MST algorithm for cable systems for which ILP is not able to find the optimal solution within acceptable time.

\subsection{ILP Formulation}
\label{secsub:ILP Algorithm Description}
Our ILP-based method is derived from Martin's formulation which expresses the MST problem in terms of a number of polynomial constraints~\cite{martin1991using,dara2019integer}. For the graph $G=(V,E)$, the formulation is given below.
\begin{mini!}
	{x_{ij}}
	{\sum_{(i, j) \in E} c_{i j} x_{i j},\label{eq:MSTlpmin}}
	{\label{eq:MSTlpmin13}}{}
	\addConstraint{\sum_{(i, j) \in E} x_{i j}}{= n-1, \label{eq:MSTlpst1}}
	\addConstraint{y_{i j}^{k}+y_{j i}^{k}}{= x_{i j},  \label{eq:MSTlpst2} \quad \forall(i, j) \in E,\  k \in V}
	\addConstraint{\sum_{k \in V \backslash\{j\}} y_{i k}^{j}+x_{i j}}{= 1, \label{eq:MSTlpst3} \quad \forall(i, j) \in E,  \ k \in V}
	\addConstraint{x_{ij}, y_{ij}^{k} \in\{0,1\}. \label{eq:MSTlpmin4}}
\end{mini!}
Here, $c_{ij}$ represents the cost of the edge $(i,j)$. The variables $x_{ij}$ and $y_{ij}^{k}$ are all binary, where $x_{ij} =1$ indicates that the edge $(i,j)$ is included in the spanning tree. The statement $y_{ij}^{k} =1$ indicates that edge $(i,j)$ is in the spanning tree and node $k$ is on the side of $j$, i.e., the connection between nodes $k$ and $i$ must go through node $j$.

Constraint~(\ref{eq:MSTlpst1}) is derived from the properties required of the tree topology, and provides a guarantee  that the number of edges is one less than the number of nodes. Constraint~(\ref{eq:MSTlpst2}) for $(i,j) \in E, k \in V$ establishes that,  if $(i,j) \in E$ is chosen to be a member  of the tree (that is, $x_{ij}=1$), then any node $k \in V$ has either to be on the same side as $j$ ($y_{ij}^{k}=1$) or on the same  side as  $i$ ($y_{ji}^{k}=1$). If the edge $(i,j) \in E$ is not in the tree (i.e., $x_{ij}=0$), then  no node ($k$) is on the side of $j$ or $i$ ($y_{ij}^{k}=y_{ji}^{k}=0$). Constraint~(\ref{eq:MSTlpst3}) for an edge $(i, j) \in E$ means that, if $(i, j) \in E$ is in the tree ($x_{i j}=1$),  edges $(i, k)$ connecting to $i$ have to be on the same side as $i$. If the edge $(i, j) \in E$ is not in the tree ($x_{i j}=0$), then some edge $(i,k)$ must exist so that node $j$ is connected to node $i$ through node $k$ (i.e., $y_{i k}^{j}=1$).

The use of Martin's formulation for the MST problem allows us to add the following inequalities~(\ref{eq:inpath}) and~(\ref{eq:inpathhop}) to enforce the cable length and hops (i.e., the number of intermediate nodes minus one) constraints (\ref{eq:PMSTlpst1}), (\ref{eq:PMSTlpst2}), respectively. For any $(a \leftrightarrow b) \in \mathbb{C}$, we have
\begin{align}
\label{eq:inpath}
\sum_{(i, j) \in E}\left(y_{ij}^{a} \cdot y_{ji}^{b}+y_{ij}^{b} \cdot y_{ji}^{a}\right) {{l_{i,j}}} &\leq l_{a,b}^*,\\
\label{eq:inpathhop}
\sum_{(i, j) \in E}\left(y_{ij}^{a} \cdot y_{ji}^{b}+y_{ij}^{b} \cdot y_{ji}^{a}\right)  &\leq N_{a,b}^*.
\end{align}
As in the previous definition, $y_{i j}^{a}=1$ means that edge $(i, j)$ is in the spanning tree and node $a$ is on the side of node $j$, $y_{ji}^{b}=1$ means that edge $(i, j)$ is in the spanning tree and node $b$ is on the side of $i$. If both $y_{i j}^{a}=1$ and $y_{j i}^{b}=1$ then  the edge $(i,j)$ is included in the spanning tree and, moreover, in the path between node $a$ and node $b$. In this case, node $a$ is close to node $j$ and node $b$ is close to node $i$ which gives  a direction to edge $(i,j)$. However, the cable path between node $a$ and node $b$ does not have a preferred direction, so we need  also to  analyze $y_{ji}^{a}$ and $y_{ij}^{b}$ in the same way. The four variables $y_{i j}^{a}$, $y_{ji}^{b}$, $y_{ji}^{a}$ and $y_{ij}^{b}$ decide whether the edge $(i,j)$ is included in the path between $a$ and $b$, without consideration of  the direction. Recall that $N_{a,b}^*$ in (\ref{eq:inpathhop}) is the maximum number of hops between node $a$ and node $b$. If the edge $(i,j)$ is included in the path between nodes $a$ and $b$, the number of hops increases by one.

Unfortunately, (\ref{eq:inpath}) and (\ref{eq:inpathhop}) are non-linear. To apply ILP techniques, we need to replace them by linear constraints. To this end, we introduce two new variables $z_{ij}^{ab}$ and $z_{ji}^{ab}$, and make $z_{ij}^{ab}= y_{ij}^{a} \cdot y_{ji}^{b}$ and $z_{ji}^{ab}= y_{ji}^{a} \cdot y_{ij}^{b}$. This means that $z_{ij}^{ab}=1$ only when both $y_{ij}^{a}=1$ and $y_{ji}^{b}=1$; otherwise, $z_{ij}^{ab}=0$. Because $y_{ij}^{a}$, $y_{ji}^{b}$ and $z_{ij}^{ab}$ are binary variables representing Boolean values, the relationship between them can be regarded as a Boolean operation. So we can use four linear constraints to express a single Boolean constraint, as~(\ref{eq:zlinear}) shows. The analysis is the same for $z_{ji}^{ab}$.
\begin{equation}
\label{eq:zlinear}
    \left\{\begin{array}{l}
z_{ij}^{ab} \leqslant y_{ij}^{a}+y_{ij}^{b}, \\
z_{ij}^{ab} \geqslant y_{ij}^{a}+y_{ij}^{b}-1, \\
z_{ij}^{ab} \leqslant 1-y_{ij}^{a}+y_{ij}^{b}, \\
z_{ij}^{ab} \leq 1+y_{ij}^{a}-y_{ij}^{b}.
\end{array}\right.
\end{equation}
Our problem is then reformulated as:
\begin{mini!}
	{x_{ij}}
	{\sum_{(i, j) \in E} c_{i j} x_{i j},\label{eq:MSTlpminbound}}
	{\label{eq:MSTlpminbound15}}{}
	\addConstraint{\sum_{(i, j) \in E} x_{i j}}{= n-1, \label{eq:MSTlpstbound1}}
	\addConstraint{y_{i j}^{k}+y_{j i}^{k}}{= x_{i j},  \label{eq:MSTlpstbound2} \quad \forall(i, j) \in E,\  k \in V}
	\addConstraint{\sum_{k \in V \backslash\{j\}} y_{i k}^{j}+x_{i j}}{= 1, \label{eq:MSTlpstbound3} \quad \forall(i, j) \in E,  \ k \in V}
	\addConstraint{\sum_{(i, j) \in E} (z_{ij}^{ab}+z_{ji}^{ab}) \cdot l_{i,j}}{\leq l_{a,b}^*,\label{eq:MSTlpstbound4}  \quad \forall (a\leftrightarrow b) \in \mathbb{C}}
	\addConstraint{\sum_{(i, j) \in E} (z_{ij}^{ab}+z_{ji}^{ab})}{\leq N_{a,b}^*, \label{eq:MSTlpstbound5} \quad \forall (a \leftrightarrow b) \in \mathbb{C}}
	\addConstraint{x_{ij}, y_{ij}^{k}, z_{ij}^{ab} \in \{0,1\}. \label{eq:MSTlpmin4}}
\end{mini!}

Constraints~(\ref{eq:MSTlpstbound1})-(\ref{eq:MSTlpstbound3}) are the same as constraints~(\ref{eq:MSTlpst1})-(\ref{eq:MSTlpst3}) and enforce the tree topology. Constraint~(\ref{eq:MSTlpstbound4}) is used for the  length constraint and constraint~(\ref{eq:MSTlpstbound5}) is used for the constraint on the number of hops.

The variables in the ILP-based formulation include $x_{ij}$, $y_{ij}^k$ and $z_{ij}$. For a complete graph, $|E|=  1/2 \cdot |V|\cdot(|V|-1)$. For MST with one pair of latency constraint~(i.e., $|\mathbb{C}| = 1$), the number of variables is $1/2 \cdot(|V|^3-|V|)$. More generally, we analyze the complexity in the case of an incomplete but connected graph. In this case, the total number of variables in the formulation is the sum of the number of $x$ (=$|E|$), the number of $y$ (=$|E|\cdot (|V|-2)$), and the number of $z$ (=$2|E||\mathbb{C}|$). The number of constraints in the ILP-based formulation is one plus the sum of two times the number of $x$ (=$2|E|$), according to ~(\ref{eq:MSTlpstbound1})-(\ref{eq:MSTlpstbound3}), and plus nine times the number of latency constraints requirements (=$9|\mathbb{C}|$), according to~(\ref{eq:zlinear}) and~(\ref{eq:MSTlpstbound4})-(\ref{eq:MSTlpstbound5}).

 Solution of the ILP formulation above provides a solution of the MST problem. To obtain numerical results for this ILP problem, we use the package python-pulp~\cite{mitchell2011pulp} that employs a method called ``branch-and-cut'',  an exact algorithm based on a combination of the branch-and-bound algorithm and the cutting plane method.

\subsection{Computational Complexity Analysis}
\label{sec:Complexity analysis of ILP algorithm}
Much research has been devoted to determining the computational complexity of ILP problems. Kannan {\it et al.}~\cite{kannan1978computational} showed that ILP problems are all NP-hard. More specifically, 0-1 integer linear programming, a  special case of the general ILP problem, is one of Karp's well-known 21 NP-complete problems~\cite{karp1972reducibility}.
In this paper, we consider an extension of the MST problem. In addition to the well studied problem of finding an MST in a weighted, undirected connected graph, there exist constraints (length or intermediate nodes) between pairs of nodes. Through the above statement, we can effectively explain the MST with constraints problem solved by our ILP formulation is NP-complete.
In~\cite{dara2019integer}, a detailed computational analysis of the different ILP formulations was described in terms of run-time. We adopt a similar approach here, and illustrate the computational complexity of our ILP-based algorithm via run-times with graphs with various number of nodes and constraints.

\begin{figure}[!htp]
	\centering
	\includegraphics[width = 1.05\columnwidth]{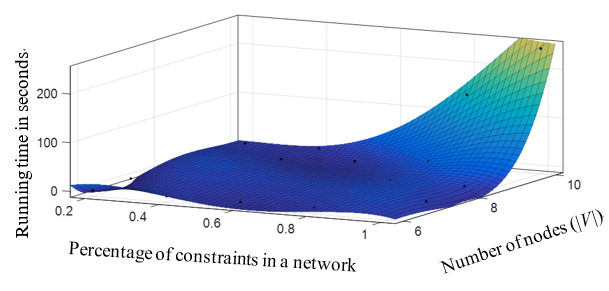}
	\caption{Run-time in seconds of ILP on partially and fully constraints network.}
    \label{fig:expf1}
\end{figure}

\begin{figure}[!htp]
	\centering
	\includegraphics[width = 0.9\columnwidth]{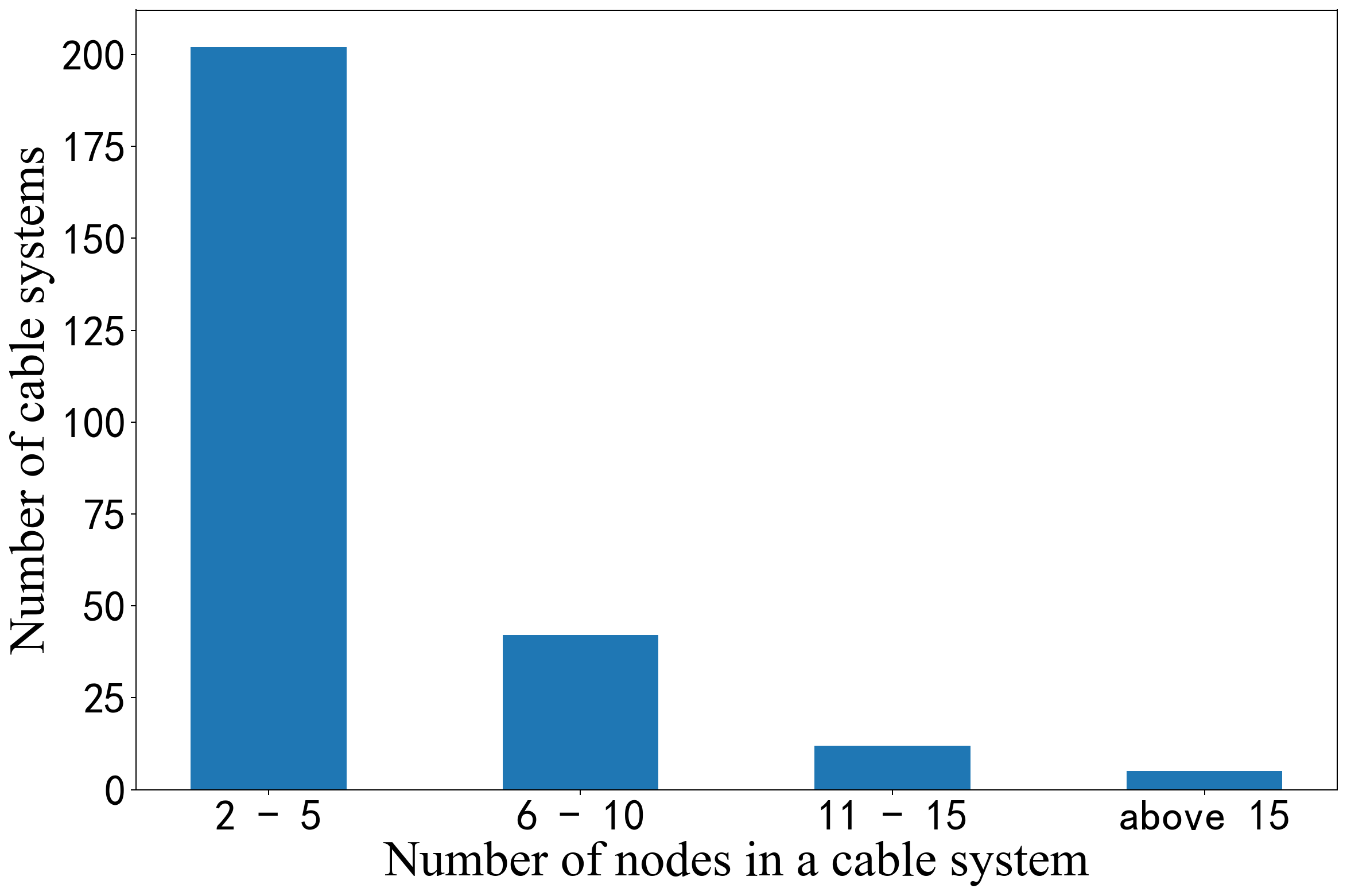}
	\caption{Histogram of the number of nodes in a cable system.}
    \label{fig:distribution}
\end{figure}

In Figure~\ref{fig:expf1}, we provide run-time of our ILP algorithm as a function of the number of nodes and constraints. This figure demonstrates that the run-time of the ILP is exponentially increasing with the number of nodes and constraints. Nevertheless, the ILP solution is applicable to most existing cable systems because the number of nodes in such systems is normally not too large, which implies that the number of latency constraints is also not too large.  This is evidenced by information available in the Submarine Cable Almanac of 2020~\cite{SubmarineTeleForum2020}. According to this information, $93\%$ the cable systems have less than 10 nodes. In Figure~\ref{fig:distribution}, we provide a histogram based on data from~\cite{SubmarineTeleForum2020} of the distribution of the number of nodes in each submarine cable systems. We note that the highest number of nodes for any existing cable system is 39 which is the number of nodes in the so-called {\it South East Asia-Middle East-Western Europe 3 cable system. }  Therefore, our ILP solution is applicable to a realistic size cable system. 

\subsection{A Heuristic Algorithm for the MST Problem With Constraints}
\label{sec:A Heuristic Algorithm for MST Problem With Constraints}
Compared with our ILP-based algorithm, the heuristic algorithm that we propose, a modification of Prim's algorithm, can find a small spanning tree that satisfies the constraints if the problem is feasible, but optimality is not  guaranteed.
We consider a fully connected network modeled as a graph $G = (V,E)$. The weight of each edge is the cable cost (which is linear to the cable length) between its end-nodes. At each step of Prim's algorithm, we first check that all latency constraints are not violated. If this is the case, for all steps, the final solution is obtained by Prim's algorithm. Otherwise, if the addition of an edge to the tree causes a violation of a latency constraint we choose the next least-cost edge to be added to the tree. The algorithm continues until either a feasible solution is obtained or the problem is infeasible. See Algorithm~\ref{alg:1}.

As in Prim's algorithm, finding the next closest node (Line 4 in Algorithm~\ref{alg:1}) takes $\mathrm{O}(|V|)$ executions and the while-loop from Line~3 to Line~11 requires $\mathrm{O}(|V|)$ executions. The do-loop from Line~5 to
Line~11 requires $|\mathbb{C}| \cdot (|V|-1)$ executions to update the path length from the start/end points in $\mathbb{C}$. The total complexity of this modification of Prim's algorithm with latency constraints is $\mathrm{O}(|V|\cdot (|V|+1) + |\mathbb{C}| \cdot (|V|+1))$ which is equal to $\mathrm{O}(|V|^2)$.

\begin{algorithm}[htb]
\caption{Prim-based method for MST with constraints.}
\label{alg:1}
\begin{algorithmic}[1]
\REQUIRE~~\\
 The graph $G = (V, E)$, and the set of constraints requirements $\mathbb{C}$.
\ENSURE ~~\\ %
A spanning tree $T$.

\STATE Let $U = \{i\}$, $i$, an arbitrary node in $V$;\\
\STATE Let $F = \emptyset$,  used to store edges in the spanning tree;\\
\WHILE{$U!=V$}
\STATE Find the smallest edge $(i,j)$, where the node $i \in U$ and $j \in V \setminus U$. \label{code:find}\\
\IF{$T = (U, F \cup \{i,j\})$ satisfies $\mathbb{C}$,}
\STATE $U = U \cup \{j\}$ and $F = F \cup (i,j)$;
\STATE Return to Step 3;
\ELSE
\STATE Eliminate this edge from $E$, and return  to Step 3;
\ENDIF
\ENDWHILE
\RETURN $T$;
\end{algorithmic}
\end{algorithm}

The advantage of the Prim-based algorithm is that it can be applied in large-scale cable network optimization where there may be many nodes. We apply the heuristic algorithm in a graph with 100 nodes to find the MST that satisfies the length constraints. The positions of these 100 nodes are randomly generated in the region $[0,500] \times [0,500]$. The length of the edge connecting two nodes is defined as the Euclidean distance between them. The constraint requirement is that the distance between nodes 0 and 50 should be less than $300$. The resultant spanning tree satisfying the constraint is shown in Figure~\ref{fig:area}.
\begin{figure}[!htp]
	\centering
	\includegraphics[width = 0.9\columnwidth]{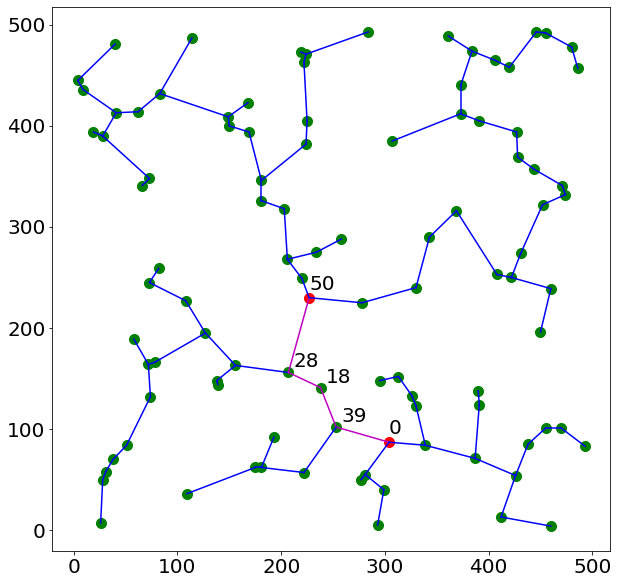}
	\caption{The spanning tree satisfying the constraint obtained by the Prim-based algorithm.}
	\label{fig:prim100nodes}
\end{figure}

The path between nodes 0 and 50 includes $0--39, 39--18, 18--28, 28--50$, of which the total length is $205.99$. The total length of the resultant spanning tree is $3289.37$.
The CPU running time is less than 0.1 ms. Note that the Prim-based algorithm cannot guarantee the spanning tree is optimal but can ensure the spanning tree satisfies the constraints while its total length is within reasonable bounds.

\section{Application of the ILP-based Algorithm}
\label{sec:Application of ILP Algorithm}
\begin{figure}[!htp]
	\centering
	\includegraphics[width = 0.9\columnwidth]{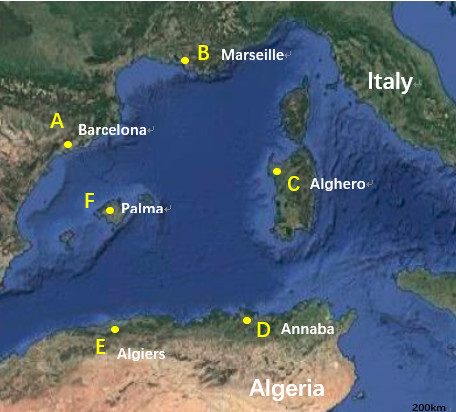}
	\caption{Region $\mathbb{D}$. Source: Google Earth.}
	\label{fig:area}
\end{figure}

\begin{figure}[!htp]
	\centering
	\includegraphics[width = 0.9\columnwidth]{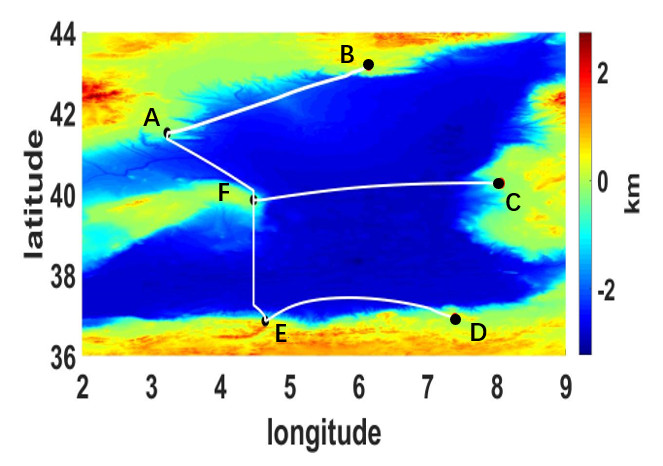}
	\caption{Minimum cost cable network without constraints.}
	\label{fig:WNOCONRISK}
\end{figure}

\begin{figure}[!htp]
	\centering
	\includegraphics[width = 1\columnwidth]{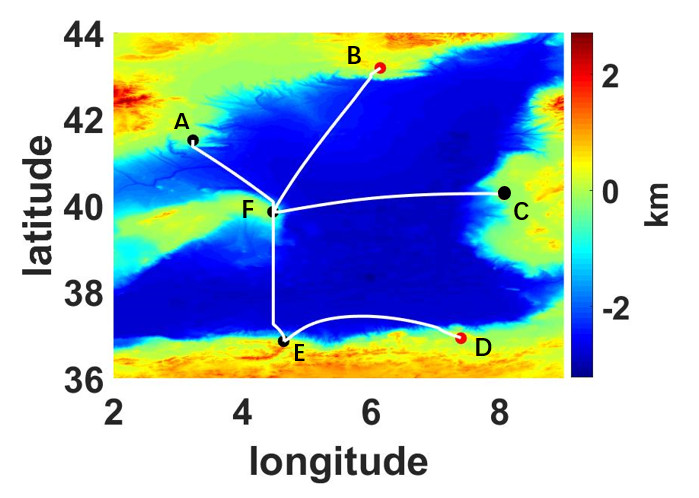}
	\caption{Minimum cost cable network with  latency constraints between B and D.}
	\label{fig:wcon1}
\end{figure}

\begin{figure}[!htp]
	\centering
	\includegraphics[width = 0.9\columnwidth]{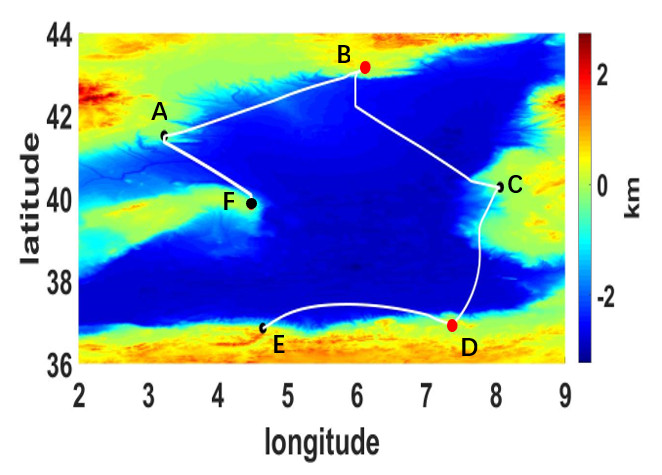}
	\caption{Minimum cost cable network with  tighter  latency constraints between B and D.}
	\label{fig:wcon2}
\end{figure}

In this section, we apply the ILP-based algorithm to a 3D realistic scenario. We use bathymetric data from the Global Multi-Resolution Topography synthesis~\cite{GMRT}.
The object region $\mathbb{D}$ is from a northwest corner $(44.000\degree \mathrm{N},2.000\degree \mathrm{E})$ to the southeast corner $(36.000\degree \mathrm{N},9.000\degree \mathrm{E})$, as shown in Figure~\ref{fig:area}. We plan a submarine cable communication network using a spanning tree topology between these six cities: Barcelona $\left(41.386\degree \mathrm{N}, 2.190\degree \mathrm{E}\right)$, Marseille $\left(43.297\degree \mathrm{N}, 5.359\degree \mathrm{E}\right)$, Alghero $\left(40.557\degree \mathrm{N}, 8.312\degree \mathrm{E}\right)$, Annaba $\left(36.928\degree \mathrm{N}, 7.760\degree \mathrm{E}\right)$, Algiers $\left(36.761\degree \mathrm{N}, 3.074\degree \mathrm{E}\right)$ and Palma $\left(39.576\degree \mathrm{N}, 2.632\degree \mathrm{E}\right)$ denoted as A, B, C, D, E, F, in Figures~\ref{fig:area}-\ref{fig:wcon2}, respectively.
We assume these six locations and the available links between them form a complete graph. Knowing the cost and length of each edge, we can use our ILP-based algorithm to construct a submarine cable network with minimal cost satisfying the given constraints. Firstly, we use FMM -- length only to find the optimal cable path for every pair of nodes. In order to compare the difference between, on the one hand,  the optimal cable paths taking account of risk and, we run FMM -- length and other considerations again for every pair of nodes to find the optimal path with minimal cost. In addition, we also calculate the great-circle distance between pairs of nodes using their geographic coordinate which are the length of smooth curves. The length and cost of these cable edges calculated by different approaches are recorded in table~\ref{tab:compare}. The result of MST for submarine cable network based on the result of FMM -- length and other considerations is shown in Figure~\ref{fig:WNOCONRISK}. Here, we suppose there are no constraints between any pairs of nodes. The result of MST by the other two approaches are same. From the comparison of the records in the table, the difference among FMM -- length and other considerations, FMM -- length only and  great-circle distance is small. However, FMM -- length and other considerations and FMM -- length only are more close to realistic result, and in this area, the consideration of the risk in cable contribution cost makes a small contribution. Accordingly, in this area,  for simplify, we can use FMM -- length only to find the optimal cable path.
\begin{table*}[!t]
\renewcommand{\arraystretch}{1.3}
\caption{Length and cost of edges between pairs of nodes.}
\label{tab:compare}
\centering
\begin{tabular}{|c|c|c|c|c|c|c|}
\hline \multirow{2}{*} {Edges} & \multicolumn{2}{c|} {FMM -- length and other considerations } & \multicolumn{2}{c|} { FMM -- length only } & \multicolumn{2}{c|} {great-circle distance } \\
\cline { 2 - 7 } & Length (km) & cost (million \$) & Length (km) & cost (million \$) & Length (km) & cost (million \$) \\
\hline AB & 304.11 & 8515.86 & 303.96 & 8511.12 & 302.75& 8480.11 \\
\hline AC & 440.44 & 12328.47 & 433.62 & 12142.55 &428.95& 12015.11 \\
\hline AD & 637.73 & 1785.86 & 623.64 & 17456.69 & 622.96 & 17449.34 \\
\hline AE & 651.82 & 15451.18 & 653.64 & 15602.13 & 533.02 & 14930.04 \\
\hline AF & 216.04 & 6049.20 & 219.51 & 6146.24 & 213.14 & 5970.06 \\
\hline BC & 379.01 & 10613.87 & 373.53 & 10459.96 & 361.13 & 10115.40 \\
\hline BD & 718.34 & 20105.46 & 727.92 & 20382.83 & 703.02 & 19691.71 \\
\hline BE & 715.08 & 20022.31 & 715.15 & 20024.38 & 715.66 & 20045.79 \\
\hline BF & 395.19 & 11084.32 & 395.64 & 11078.96 & 395.69 & 11083.49 \\
\hline CD & 379.45 & 10625.66 & 375.27 & 10508.75 & 375.13 & 10507.55 \\
\hline CE & 483.41 & 13536.97 & 482.06 & 13498.93 & 482.06 & 13501.63 \\
\hline CF & 310.76 & 8701.67 & 309.71 & 8671.92 & 308.16 & 8631.69 \\
\hline DE & 281.95 & 7875.72 & 245.53 & 6875.06 & 244.52 & 6849.17 \\
\hline DF & 446.62 & 12506.61 & 412.89 & 11561.92 & 411.87 & 11536.59 \\
\hline EF & 337.17 & 9440.81 & 337.60 & 9452.99 & 333.90 & 9352.55 \\
\hline
\end{tabular}
\end{table*}

As discussed in Section~\ref{sec:Introduction}, submarine cable network construction often needs to take account of the latency constraints for communication between different pairs of nodes. In order to clearly show our method works well on the constraints consideration, we assume that nodes B and D have a strict latency requirement that the cable length between them is less than 1100 km ($d=1100$), the number of hops is no more than three ($h=3$), and there is no latency requirements of other pairs of nodes. In the previously generated MST, the cable length between nodes B and D is the sum of edges length between nodes BA, AF, FE and ED (totally, $1107$). We add two constraints here to achieve the requirement. In accord with~(\ref{eq:MSTlpstbound}), the two constraints are as shown below.
\begin{equation}
\label{eq:MSTlpstbound}
\left\{\begin{array}{l}
\sum_{(i, j) \in E} (z_{ij}^{\text{BD}}+z_{ji}^{\text{BD}})  l_{i,j} \leq 1100,\\\\
\sum_{(i, j) \in E} (z_{ij}^{\text{BD}}+z_{ji}^{\text{BD}})  \leq 3,
\end{array}\right.
\end{equation}
where $E$ is the set of all edges between pairs of nodes, $i$,$j$ $\in\{\text{A},\text{B},\text{C},\text{D},\text{E},\text{F}\}$.
Figure~\ref{fig:wcon1} shows the result of our ILP-based algorithm for finding MST for the submarine network with the constraints ($d=1100$, $h=3$) between nodes B and D.

To show the flexibility of our algorithm, we make another example finding MST for the submarine network with tighter latency constraints ($d=800$, $h=2$) between nodes B and D. Figure~\ref{fig:wcon2} shows the result of our ILP-based algorithm for finding MST for the submarine network with tighter latency constraints between nodes B and D.

\section{Conclusion}
\label{sec:conclusion}
We have provided a method, called FMM/ILP, for optimizing a tree-topology cable network with latency constraints in an irregular 2D manifold in a 3D Euclidean space. Specifically, in the submarine cable application,  while focusing on minimizing the cost of the entire network, we ensure that cable length and number of hops between any pair of nodes satisfy latency requirements. Our FMM/ILP method is based on finding a cable path and cost between any pair of nodes using FMM and then finding an MST with latency constraints (i.e. constraints on cable length and the number of hops) for any pair of nodes, based on ILP.

We have proposed a new ILP method based on Martin's formulation to solve this MST with constraints. Although, in general, ILP is not scalable, we have shown that it can find an MST with latency constraints for most realistic cable systems in a few minutes.  An alternative heuristic algorithm based on the application of Prim's algorithm to finding an MST with constraints has been demonstrated to provide approximate solutions for large scale cable system.

A limitation is that we have made a simplifying assumption that the cable cost is linear to its length. Such an assumption is applicable to many cable systems around the world; in particular,  we have demonstrated the validity of this assumption for a potential cable system connecting six cities in the Mediterranean Sea. For this validation, we compared the path obtained by the approaches we call ``FMM -- length only'' and  ``FMM -- length and other considerations'' and have observed only small differences in the resulting optimal paths. Consistent results have also been obtained for paths and their costs with a third approach based on the great-circle distance. Using optimal cable routing based on FMM -- length only between all pairs of nodes, we applied an ILP-based algorithm to find an MST with latency constraints. Finally, we have applied our FMM/ILP-based algorithm to the cable system example in the Mediterranean and have demonstrated that it can find the MST with latency constraints between pairs of nodes.

\bibliographystyle{IEEEtran}
\bibliography{main}
\end{document}